\documentclass[11pt]{article}
\usepackage{palatino}
\usepackage{fullpage,appendix}
\usepackage{enumerate,color,graphicx}
\usepackage{array}
\usepackage[fleqn]{amsmath}
\usepackage{amssymb,amsthm,amsfonts}
\usepackage{hyperref}
\usepackage{subfigure}

\begin{document}

\title{Pushing Your Point of View: Behavioral Measures of Manipulation
  in Wikipedia}

\author{ Sanmay Das, Allen Lavoie, and Malik Magdon-Ismail\\
       Dept. of Computer Science\\
       Rensselaer Polytechnic Institute\\
       Troy, NY 12180\\
}

\maketitle
\begin{abstract}

As a major source for information on virtually any topic, Wikipedia
serves an important role in public dissemination and consumption of
knowledge. As a result, it presents tremendous potential for people to
promulgate their own points of view; such efforts may be more
subtle than typical vandalism. 
In this paper, we introduce new behavioral metrics to quantify the
level of controversy associated with a particular user: a Controversy
Score (C-Score) based on the amount of attention the user focuses on
controversial pages, and a Clustered Controversy Score (CC-Score) that
also takes into account topical clustering. We show that both these
measures are useful for identifying people who try to ``push'' their
points of view, by showing that they are good predictors of which
editors get blocked. The metrics can be used to triage potential POV
pushers. We apply this idea to a dataset of users who requested
promotion to administrator status and easily identify some editors who
significantly changed their behavior upon becoming administrators. At
the same time, such behavior is not rampant. Those who are promoted to
administrator status tend to have more stable behavior than comparable
groups of prolific editors. This suggests that the Adminship process
works well, and that the Wikipedia community is not overwhelmed by
users who become administrators to promote their own points of view.

\end{abstract}

\section{Introduction}
\label{sec:intro}
Wikipedia has become a one-stop source for information on nearly any
subject. In aggregate, it has the power to broadly influence public
perceptions. Wikipedia's ubiquity creates strong incentives for biased
editing, attracting editors with strong opinions on controversial
topics. At the same time, Wikipedia is self-policing, and over time
the Wikipedia community has formulated a comprehensive set of policies
to prevent editors from ``pushing'' their own points of view on the
readership (``POV pushing''). Most policing takes the form of users
editing or reverting disputed content, while persistent violations are
brought to the attention of administrators. A responding administrator
has the power to temporarily protect pages from being edited, and to
block users from editing. POV pushing is not considered vandalism on
Wikipedia, the latter term being reserved for blatantly
ill-intentioned edits.

While there has been a lot of attention paid to the problems of
vandalism and maintenance on Wikipedia, there has been little
systematic, quantitative investigation of the phenomenon of POV
pushing.  Nevertheless, anecdotal evidence suggests that it is a
serious issue. For example, in April 2008 a pro-Palestinian online
publication called Electronic Intifada released messages from the
pro-Israel media watchdog group CAMERA (the Committee for Accuracy in
Middle East Reporting in America) that asked for volunteers to help
``keep Israel related entries ... from becoming tainted by anti-Israel
editors.''  The messages also contained blueprints explaining how
members could become Wikipedia administrators, and then use their
power to further the goals of the organization \cite{Candid2008}. In
2006, there was a significant controversy surrounding edits to
Wikipedia pages of prominent U.S.\ politicians, made by their own
staff.\footnote{%
\url{http://en.wikipedia.org/wiki/Congressional_Staffer_Edits} retrieved 11/4/2011.
}

To further illustrate the importance of the problem, in an interview
with Alex Beam of the \emph{Boston Globe}, Gilead Ini, who initiated
the CAMERA campaign, said ``[Wikipedia] may be the most influential
source of information in the world today, and we and many others think
it is broken'' \cite{beam2008war}. But another quote from Ini
highlights the difficulty of confronting these issues, ``Wikipedia is
a madhouse. We were making a good-faith effort to ensure accuracy.''
Indeed, Wikipedia policies typically assume good faith. The policy on
vandalism states ``Even if misguided, willfully against consensus, or
disruptive, any good-faith effort to improve the encyclopedia is not
vandalism. Edit warring over content is not vandalism.''\footnote{
  \url{http://en.wikipedia.org/wiki/Wikipedia:Vandalism} retrieved
  11/4/2011.} In addition to the difficulty of arguing against good
faith when editors may simply be attempting to disseminate strongly
held beliefs, even more subtle forms of manipulation can achieve a
similar outcome. For example, a manipulative administrator may enforce
Wikipedia's Neutral Point Of View (NPOV) guidelines selectively,
reverting only edits that take a particular point of
view. Hypothetically, a manipulative admin with a conservative
viewpoint may revert all edits that seem to push a liberal viewpoint,
while leaving those that push a conservative viewpoint untouched (or
vice versa for a manipulative liberal admin). While there is nothing
technically ``wrong'' with this, it can significantly affect the
information on pages.

It is worth noting that even if there is attempted manipulation on
sensitive pages, it is restricted to a relatively small fraction of
Wikipedia. Most pages are more ``encyclopaedic'' in nature (for
example, pages on mathematics or on algorithms), and generate less
controversy than pages that deal with current events or ideologies.

All of this cries out for a useful algorithmic method of detecting
potential POV pushing behavior, and quantitative metrics that can
provide evidence and allow us to examine the behavior of such users in
more detail. In this paper we present two such metrics and then use
them to examine the behavior of the population of administrators.

The first metric, the Controversy Score (``C-Score''), measures the
proportion of energy an editor spends on controversial pages. It works
by first assigning a controversy score to each page. This score is
based on factors that have been identified as well-correlated with
controversy, including the number of revisions to an article's talk
page, the fraction of minor edits on the talk page, and the number of
times it has been protected \cite{Kittur2007HeSaysSheSays}. It is
independent of language or content, and therefore easily
generalizable. An editor's C-Score is then the mean of the controversy
scores of the pages she edits, weighted by the proportion of her
editing attention she focuses on those pages.

While the C-Score is a useful measure, it does not account for the
topical clustering of a user's edits. This is particularly important
when we use these scores to assess the behavior of administrators,
because administrators' responsibilities imply that they will spend
more time on controversial pages in general. However, we would expect
users who have strong opinions on a topic to push their POV especially
in pages related to that topic, rather than broadly across many
different controversial topics. Consider two editors A and B, with 100
edits each. They each have 25 edits on the article about the U.S.\
Republican party. Editor A's remaining edits are about Republican
legislators and Republican sponsored legislation from the past 10
years, while B's are divided between the IRA, the Catholic Church, and
Jimmy Wales. All of B's edits are controversial, but only some of A's
edits are. While B has more controversial edits, we would intuitively
consider A to be more suspicious.

To deal with editors of this form, we introduce the Clustered
Controversy Score (``CC-Score''), which takes into account the
similarity among different pages that a user has edited, in addition
to how controversial those pages are. We expect the CC-Score to be
particularly useful for triage, as it is designed to be a high recall
measure: it flags potentially manipulative users, who focus their
attention on specific topic areas that include controversial
topics. Of course, some users who have editing patterns of this form
may be acting in good faith and just have deep interests in that
topic.

Both the C-Score and the CC-score are behavioral. They do not rely in
any way on the specific text of edits, only on the patterns of editing
and interaction between editors. We demonstrate the validity of the
two scores by showing that they have predictive power in
discriminating between heavy editors who were blocked and equally
heavy editors who were not. Having validated them on an exogenous
measure, we then apply these measures in order to analyze the behavior
of administrators. We compare the editing patterns of administrators
who score highly on the C-Score and the CC-Score. The CC-Score
identifies administrators who would not have been identified by the
simple C-Score, because only some of the major pages on the topics
they edit are highly controversial. These administrators also edit a
long tail of related pages, thus influencing public perception of the
topic at large.

Finally, we use the CC-Score to test whether or not admins are
behaving in a truly manipulative manner: first becoming trusted and
attaining promotion to admin status, and then using this trusted
status to push their points of view in particular domains. To do so,
we look at changes in CC-score before and after the editor stood for
promotion (the Request for Adminship, or ``RfA'' process). While we
find several instances of potentially suspicious changes of focus, we
also note that the overall behavior of the population of editors who
become admins is better than that of two comparable populations: (1)
those who stood for election but failed; and (2) those who edited
prolifically but never stood for election to administrator status. The
population behavior is better in the sense that the variance of
changes in the CC-Score pre- and post-RfA is lower, indicating that
those who fail in their RfAs actually change their behavior more
significantly. Thus, Wikipedia admins as a population do not
misrepresent themselves in order to gain their trusted status.

\subsection{Contributions}

We introduce two new behavioral measures that indicate whether or not
a user is trying to push his or her point of view on Wikipedia
pages. These measures have predictive power on historical data: they
can determine which users were blocked for disputes related to
controversial editing. We anticipate that these measures can be used
for auditing or triage: they can flag potentially suspicious behavior
automatically for more detailed human investigation. The measures are
behavioral and general, and do not rely on specifics of text edited by
users, and are thus applicable beyond Wikipedia.

We then show how these measures can be used to discover interesting
changes in behavior, focusing on the behavioral changes of editors who
applied for promotion to administrator status on Wikipedia. While
there are instances of suspicious looking changes in behavior upon
promotion to administrator status, we find that at the
population-level, Wikipedia editors are in fact better behaved than
the population of prolific editors, in the sense that their behavior
is more stable, and does not change significantly upon
promotion. While there are specific instances that seem suspicious,
our evidence suggests that the Wikipedia adminship process works well
at the population level: there is no evidence that editors are in
general seeking promotion to adminship so they can ``push'' their
point of view on the larger population.

\section{Related work}
\label{sec:related}
There is a large literature on many different aspects of Wikipedia as
a collaborative community. It is now well-established that Wikipedia
articles are high quality \cite{Gil05} and very popular on the Web
\cite{Spo07}. The dynamics of how articles become high quality and how
information grows in collective media like Wikipedia have also
garnered some attention \cite{WilHub07,DasMag10}. While there has not
been much work on how Wikipedia itself influences public opinion on
particular topics, it is not hard to draw the analogy with search
engines like Google, which have the power to direct a huge portion of
the focus of public attention to specific pages. Hindman \emph{et al}
discuss how this can lead to a few highly ranked sites
coming to dominate political discussion on the Web
\cite{hindman2003googlearchy}. Subsequent research shows that the
combination of what users search for and what Google directs them to
may lead to more of a ``Googlocracy'' than the ``Googlearchy'' of
Hindman \emph{et al} \cite{menczer2006googlearchy}.

Our work in this paper draws directly on three major streams of
literature related to Wikipedia. These are, work on conflict and
controversy, automatic vandalism detection, and the process of
promotion to adminship status on Wikipedia.

There is a significant body of work characterizing conflict on
Wikipedia. Kittur \emph{et al} introduce new tools for studying
conflict and coordination costs in Wikipedia
\cite{Kittur2007HeSaysSheSays}. Vuong \emph{et al} characterize
controversial pages using both disputes on a page and the
relationships between articles and contributors
\cite{Vuong2008Ranking}. We use the measures identified by Kittur
\emph{et al} and Vuong \emph{et al} as a starting point for 
measuring the controversy level associated with a page. This then
feeds into our user-level C-Score and CC-Score measures. Our results
on the blocked users dataset serve as corroborating evidence for the
usefulness of these previously identified measures.

Automatic vandalism detection has been a topic of interest from both
the engineering perspective (many bots on Wikipedia automatically find
and revert vandalism), as well as from a scientific perspective. Smets
\emph{et al} report that existing bots, while useful, are ``far from
optimal'', and report on the results of a machine learning approach
for attempting to identify vandalism \cite{Smets2008Vandalism}. They
conclude that this is a very difficult problem to solve without
incorporating semantic information. While we touch on vandalism in
dealing with blocked users, we are focused on ``POV pushing'' by
extremely active users who are unlikely to engage in petty vandalism,
which is the focus of most work on automated vandalism detection.

Wikipedia administrator selection is an independently interesting
social process. Burke and Kraut study this process in detail and build
a model for which candidates will be successful once they choose to
stand for promotion and go through the Request for Adminship (RfA)
process \cite{Burke2008Mop}. The dataset of users who stand for
promotion is useful because it allows us to compare both previous and
later behavior of users who were successful and became admins and
those who did not.

Finally, we use a similarity metric for articles based on editors
which is similar to existing work on expert-based
similarity \cite{Li2011Mining}.

\section{Methodology}
\label{sec:methodology}
We begin by discussing our methodology in computing the ``simple''
Controversy Score for each user, and then describe how we can compute
a Clustered Controversy Score that captures editors who focus on
articles related to a single, controversial topic.

All data is from an April 2011 database dump of the English
Wikipedia. The term ``article'' refers to a page in Wikipedia's
article namespace along with any pages in the article talk namespace
with the same name, unless otherwise specified.

\subsection{Controversy Score}
We define the C-Score for a user as an edit-proportion-weighted
average of the level of controversy of each page. The controversy of a
page (loosely following the article-level conflict model of Kittur
\emph{et al} \cite{Kittur2007HeSaysSheSays}) is based on the number of
revisions to an article's talk page, the fraction of minor edits on an
article's talk page, and the number of times a page is ``protected'',
where editing by new or anonymous users is limited.

We scale and shift each of the three quantities above such that their
5th and 95th percentiles are equal, then take the mean. Next, we
transform this number such that the lowest values are at -5 and 1\% of
articles have scores above 0. Finally, the scores are transformed
using the logistic function $1 / (1 + e^{-t})$. This produces a
controversy score $c_k \in [0, 1]$ for each page. One alternative to
manual tuning is logistic regression, where a model is trained on a
data set reflecting some notion of controversy.

Let $p_k$ be the fraction of a user's edits on page $k$. The
controversy score for a user is then an edit-weighted average of the
page-level controversy scores:

\begin{equation}
  \label{eq:controversy}
  \mathrm{C-Score} = \sum_{k} p_k c_k
\end{equation}

We would expect this measure to be effective at finding users who edit
controversial pages. However, many Wikipedia users dedicate at least
part of their time to removing blatant vandalism, which occurs
disproportionally on controversial pages. Thus we turn to a measure
that combines topical clustering with controversy.

\subsection{Clustered Controversy Score}
We work from the hypothesis that users who concentrate their edits
have some vested interest in those articles. Going back to the example
in the Introduction, we would like to be able to detect users like A,
who focuses almost entirely on Republican politics. While A's edits
include some controversial pages, B fights vandalism broadly, and so
has exclusively controversial edits. B has the same number of edits to
the article on the U.S.\ Republican Party as A, but the rest of B's edits
are scattered across other topics. A's edits to this article are
interesting; they are topically related to A's other edits. At the
same time, edits to this article by editor B are far less interesting.

We would like to incorporate a measure of topical edit concentration
into the C-Score. In order to do so, we could define topics globally,
but this is both expensive and sensitive to parameter changes: what is
the correct granularity for a topic? Instead, we focus on a local
measure of topical concentration. Given a similarity metric between
articles, we can measure the extent to which a user's edits are
clustered.

\paragraph{Page similarity}

We base our score on a generalization of the clustering coefficient to
weighted networks with edge weights between 0 and 1
\cite{Kalna2007Clustering}. Several natural measures of page
similarity have values in this rage.

We consider pages which link to (or are linked from) the same pages as
similar, pages edited by the same users as similar, and pages in the
same categories as similar. Each page has a set of incoming links $I$,
outgoing links $\Theta$, users $U$, and categories $\Gamma$ associated
with it. To determine how similar two pages are based on one of these
sets, we divide the cardinality of the intersection of the sets from
each page by the cardinality of their union (the Jaccard
coefficient). To compute a single similarity score between two pages
$i$ and $j$, we take an average of scores for each type of set, giving
equal weight to links, users, and categories. The similarity score
$w_{ij}$ is then:
\begin{equation}
  w_{ij} = \frac{1}{6} \frac{|I_i \cap I_j|}{|I_i \cup I_j|} +
  \frac{1}{6} \frac{|\Theta_i \cap \Theta_j|}{|\Theta_i \cup
    \Theta_j|} + \frac{1}{3} \frac{|U_i \cap U_j|}{|U_i \cup U_j|} +
  \frac{1}{3} \frac{|\Gamma_i \cap \Gamma_j|}{|\Gamma_i \cup
    \Gamma_j|}
\end{equation}
\paragraph{Computing the CC-Score}
Consider a set of edits from a user. Let $N$ be the number of unique
pages in this set and $w_{ij}$ be the similarity score between pages
$i$ and $j$. We start with a generalization of the clustering
coefficient \cite{Kalna2007Clustering}. For a page $k$, define:
\begin{equation}
  \label{eq:clust}
  \mathrm{clust}(k) = \frac{\sum_{i=1}^{N} \sum_{j=1}^{N} w_{ki}
    w_{kj} w_{ij}}{\sum_{i=1}^{N} \sum_{j=1}^{N} w_{ki} w_{kj}}
\end{equation}
The clustering coefficient will be higher when other pages in the edit
set are related to $k$ and to each other. When computing the CC-Score
for the entire edit set, there are two other factors we would like to
consider: how much a user concentrated on any given page, and how
controversial that page is. Let $p_k$ be the proportion of edits on
page $k$, and $c_k$ be some measure of controversy. Then we have the
following coefficient for the edit set:
\begin{equation}
  \label{eq:cc}
  \mathrm{CC-Score} = \sum_{k=1}^{N}p_k c_k \mathrm{clust}(k)
\end{equation}
Since $\sum_{k=1}^N p_k = 1$, \eqref{eq:cc} is a weighted average. If
$c_k \in [0, 1]$, then so is \eqref{eq:cc}. Pushing raw controversy
scores through a sigmoid to produce $c_k$ ensures that this condition
holds, and also prevents outliers from unduly affecting the final
score.

There is no reason that $c_k$ must be a measure of
controversy. Instead, it can measure any property of a page which is
of interest. For example, a $c_k$ measuring how much a page relates to
global warming would yield a ranking of editors based on the extent to
which their edits concentrate on global warming. The CC-Score is a
general tool for ranking single-topic contributors based on some
property of that topic. We also compute a raw Clustering Score where
each page has the same $c_k$ -- this yields a measure of topical
clustering independent of any properties of the particular pages.

We choose a measure that combines clustering and
controversy page-wise rather than user-wise so that we do not end up
with editors who are very topically focused on uncontroversial pages
(say Flamingos), but also spend a significant fraction of their time
combating vandalism broadly across a spectrum of topics. We also note
that the only Wikipedia-specific contributions to the CC-Score are
encapsulated in the computation of $c_k$ and $w_{ij}$. The same
quantities can be computed for a wide variety of collaborative
networks. Consider email messages: $w_{ij}$ between two threads could
be based on senders and recipients, and $c_k$ based on the length of
the thread as a measure of controversy. These quantities are entirely
language independent, although we might make use of natural language
processing to improve estimates of both similarity and controversy.
\section{Evaluation}
\label{sec:evaluation}
We evaluate our metrics in several different ways. First, to establish
their validity, we examine whether the metrics provide discriminatory
power in identifying potentially manipulative users. In order to do
so, we need an independent measure of manipulation, so we focus on
users that were blocked from editing on Wikipedia, and compare them
with a similar set who were not blocked. One of the goals of our work
is to provide an objective metric for analyzing administrators, who
have gained significant status in Wikipedia. We present some detail on
the editing habits of the admins who score highest on our metrics. In
doing so, we also use our metrics to provide fresh insight into what
is controversial on Wikipedia, by analyzing the topic distribution of
edits amongst admins with high CC-Scores.

A reasonable hypothesis, suggested by the CAMERA messages discussed in
Section \ref{sec:intro} is that people who wish to seriously push
their points of view on Wikipedia may try to become admins by editing
innocuously, and then changing their behavior once they become
admins. In order to examine this hypothesis, we look at the behavior
of admins whose CC-Scores changed significantly, as well as at the
distribution of changes in the CC-Score.

\subsection{Blocked users}
\begin{figure}
  \begin{center}
  \includegraphics[width=.7\textwidth]{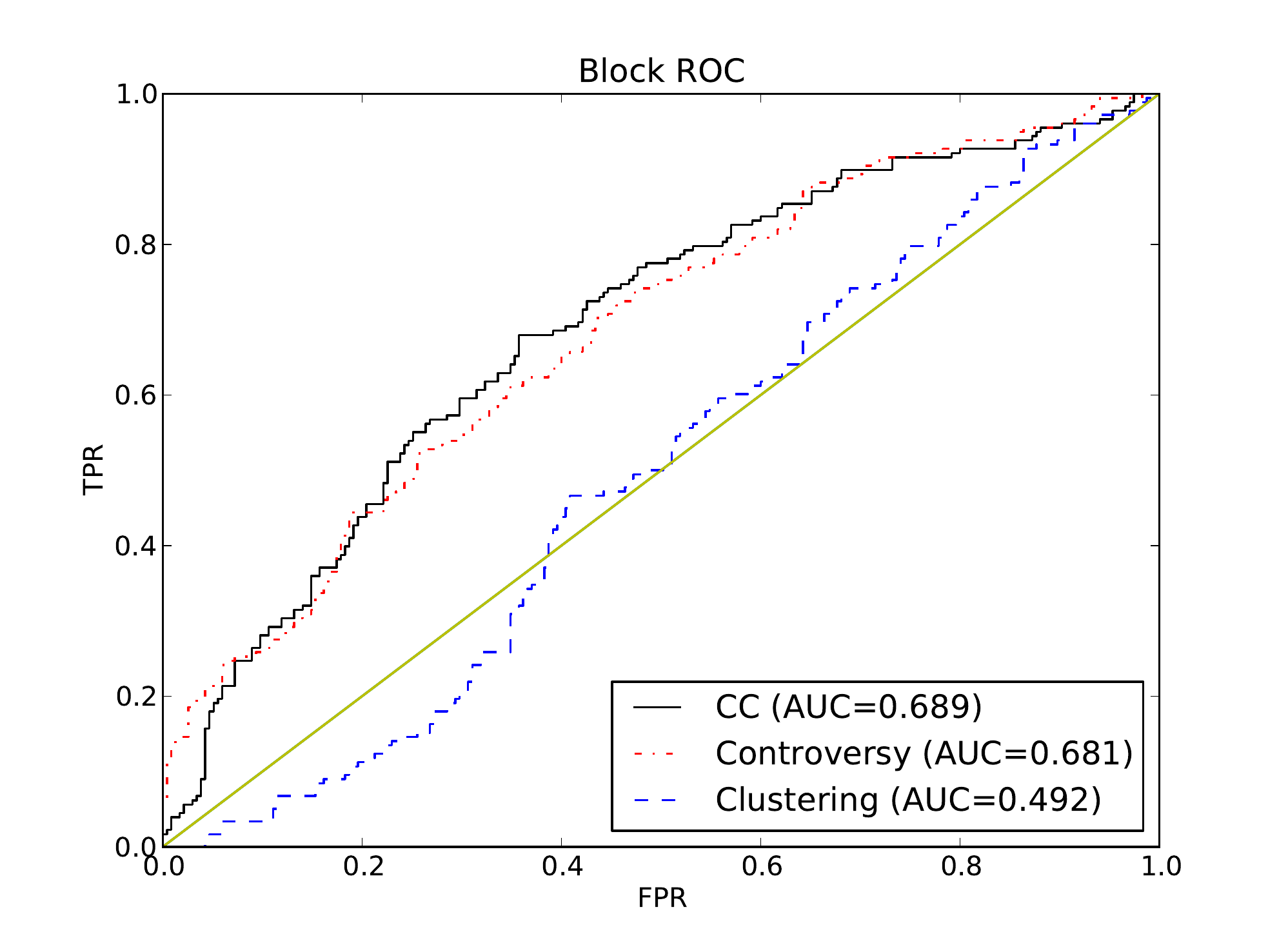} 
  \caption{ROC curve for CC, Controversy, and Clustering Scores when
    differentiating between blocked and not-blocked users, based on
    180 days of data. The CC and Controversy Scores effectively
    discriminate between these classes, whereas the Clustering Score
    does not.}  \label{figure:block_roc}
  \end{center}
\end{figure}

Users can be blocked from Wikipedia for a variety of reasons.
Reasons for blocks include blatant vandalism (erasing the content of a
page), editorial disputes (repeatedly reverting another user's edits),
threats, and more. Many blocks are of new or anonymous editors
for blatant vandalism; we are not interested in these blocks.

We are interested in blocks stemming from content disputes. While
editors are not directly blocked for contributing to controversial
articles, controversy on Wikipedia is often accompanied by ``edit
warring'', where two or more editors with mutually exclusive goals
repeatedly make changes to a page (e.g., one editor thinks the article
on Sean Hannity should be low priority for WikiProject Conservatism,
and another thinks it should be high priority).

We examine a set of users who were active between January 2010 and
April 2011. For blocked users, we use 180 days of data, directly
before the block. For the users who were never blocked, the 180 days
ends on one of their edits chosen randomly. In order to filter out new
or infrequent editors, we only consider users with between 500 and
1000 edits during this 180 day period. The upper bound removes users
who do significant amounts of automated editing; it is not uncommon
for such accounts to be blocked for misbehaving scripts which have
nothing to do with controversy. By examining only exceptionally active
users, we eliminate most petty reasons for blocks; users who have made
hundreds of legitimate contributions are unlikely to start blanking
pages. After the filtering, we are left with 178 blocked users and 236
who were never blocked.

Figure \ref{figure:block_roc} shows the performance of the CC,
Controversy, and Clustering Scores when discriminating between the
blocked users and users who were never blocked. Both the CC- and
C-Scores show significant discriminative power, while
Clustering alone is no better than guessing.

The performance of the CC- and C-Scores on the blocked users
data set validates both measures for detecting users who make
controversial contributions to Wikipedia. Many blocks in this data set
involve violations of Wikipedia's ``3 Revert Rule'', limiting the
number of contributions which an editor can revert on a single page
during any 24 hour period, which implies that editors are not only
making controversial changes but are vigorously defending them. This
rule is not automatically enforced and does not apply to blatant
vandalism; instead, another user must post a complaint which is then
reviewed by an administrator. It is certainly possible to edit
controversial pages while following the rules closely and never
getting blocked, and likewise some blocks of active users have nothing
to do with controversy (rogue scripts, for example). All else being
equal, we expect more controversial editors to be blocked more
frequently. The discriminative power of the CC- and C-Scores
provides strong evidence that these scores are correctly detecting
controversial editors.

\begin{figure}
  \begin{center}
    \includegraphics[width=0.38\textwidth]{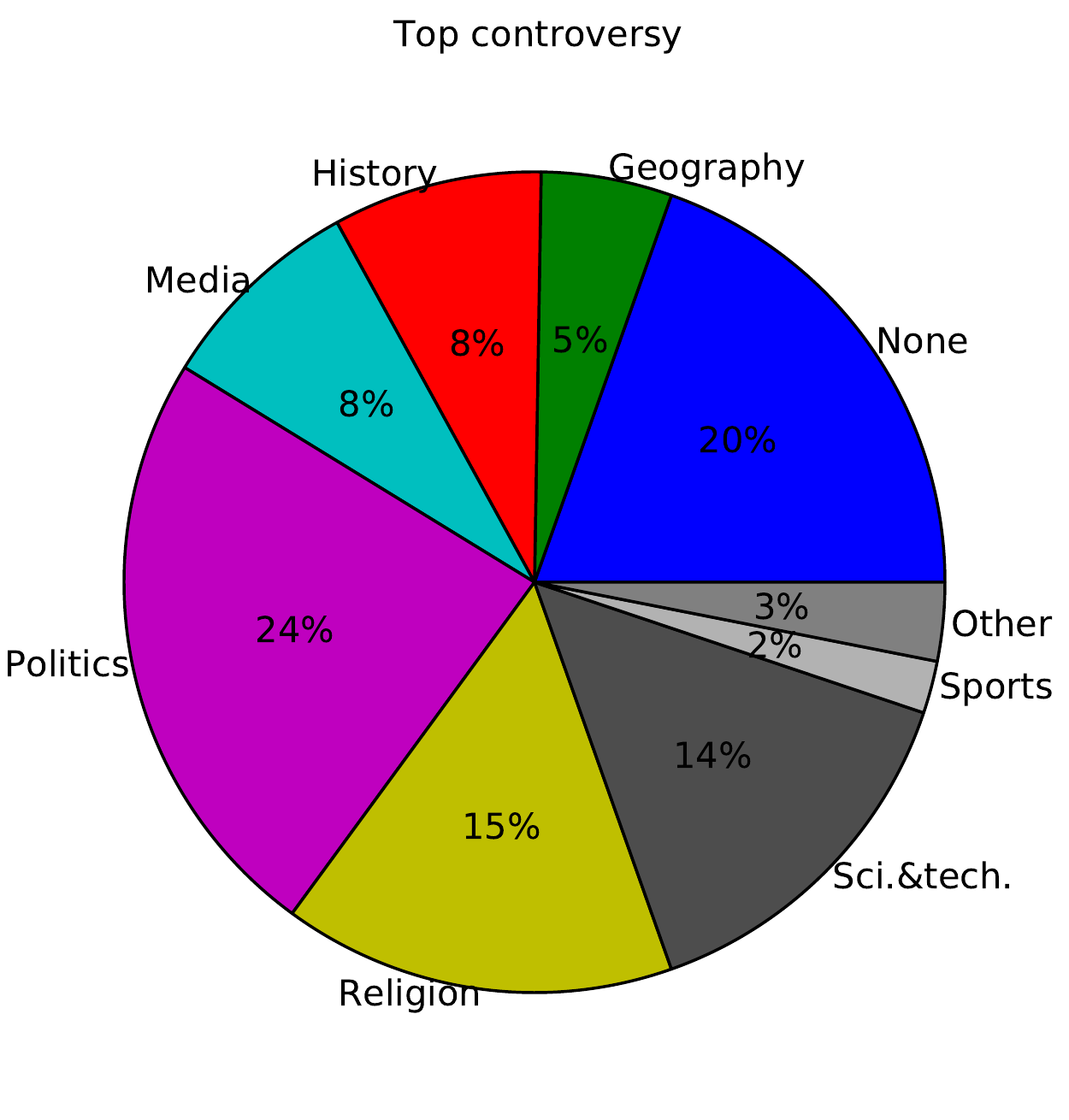}
    \includegraphics[width=0.38\textwidth]{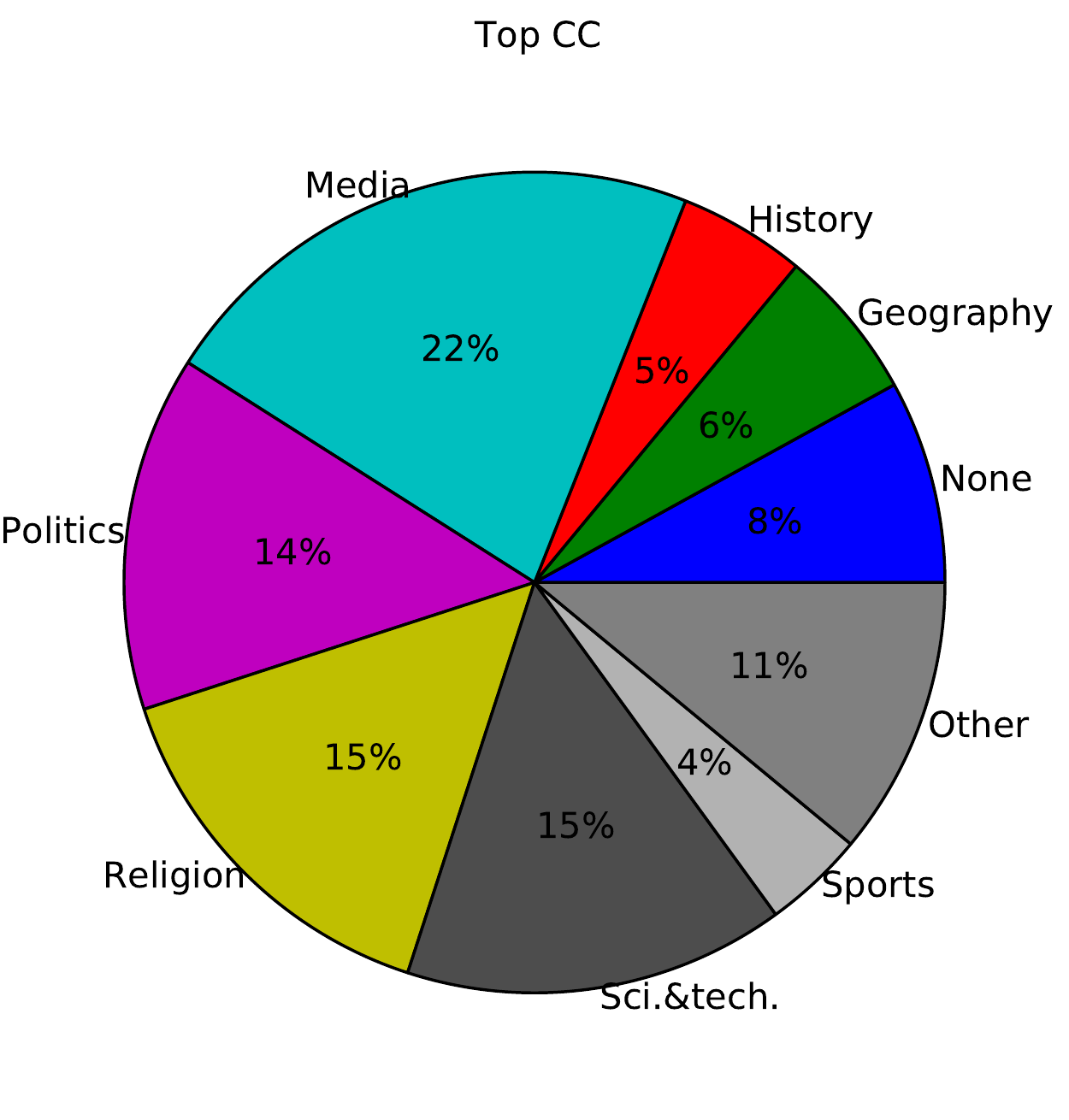}
    \includegraphics[width=0.38\textwidth]{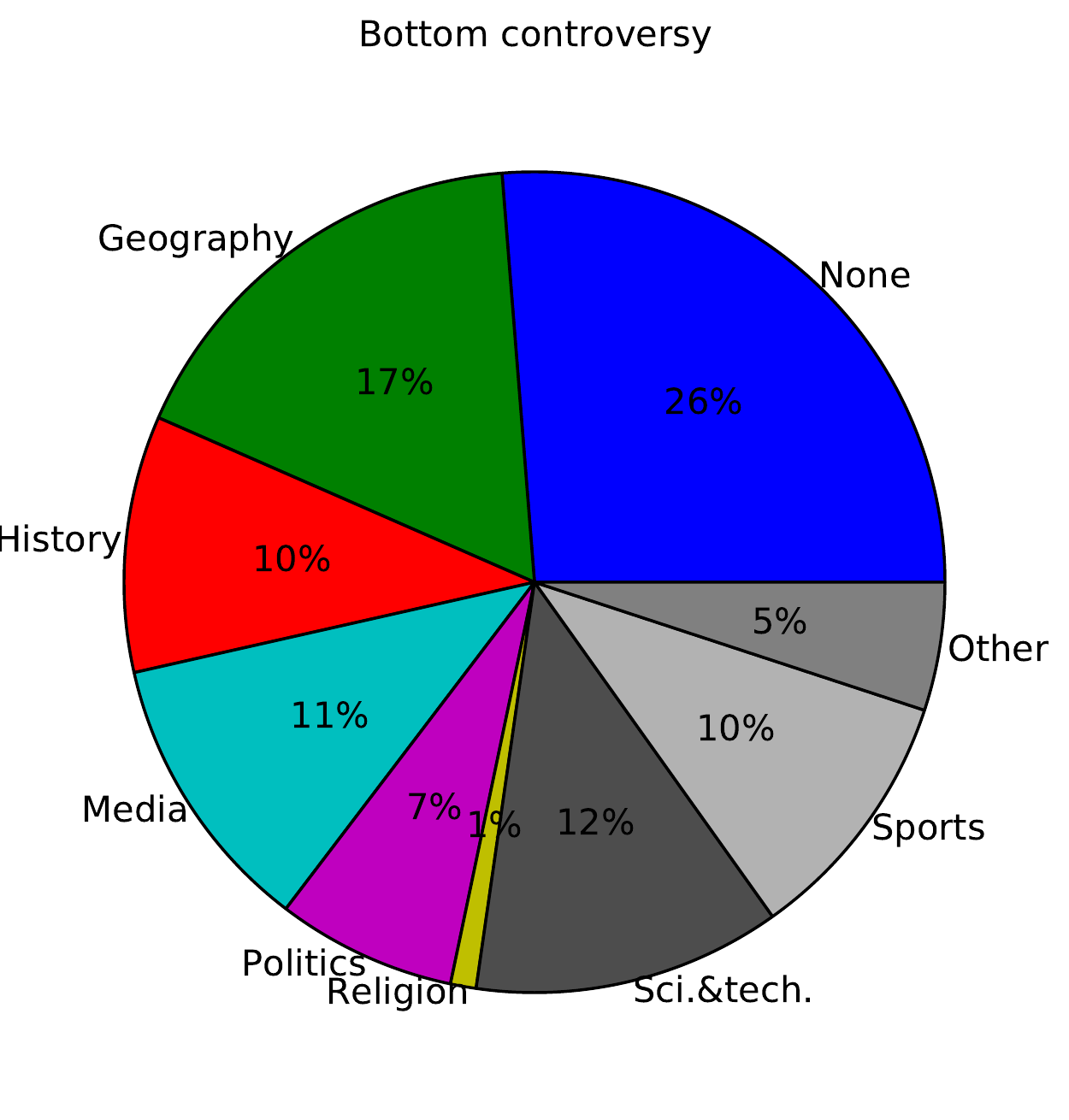}
    \includegraphics[width=0.38\textwidth]{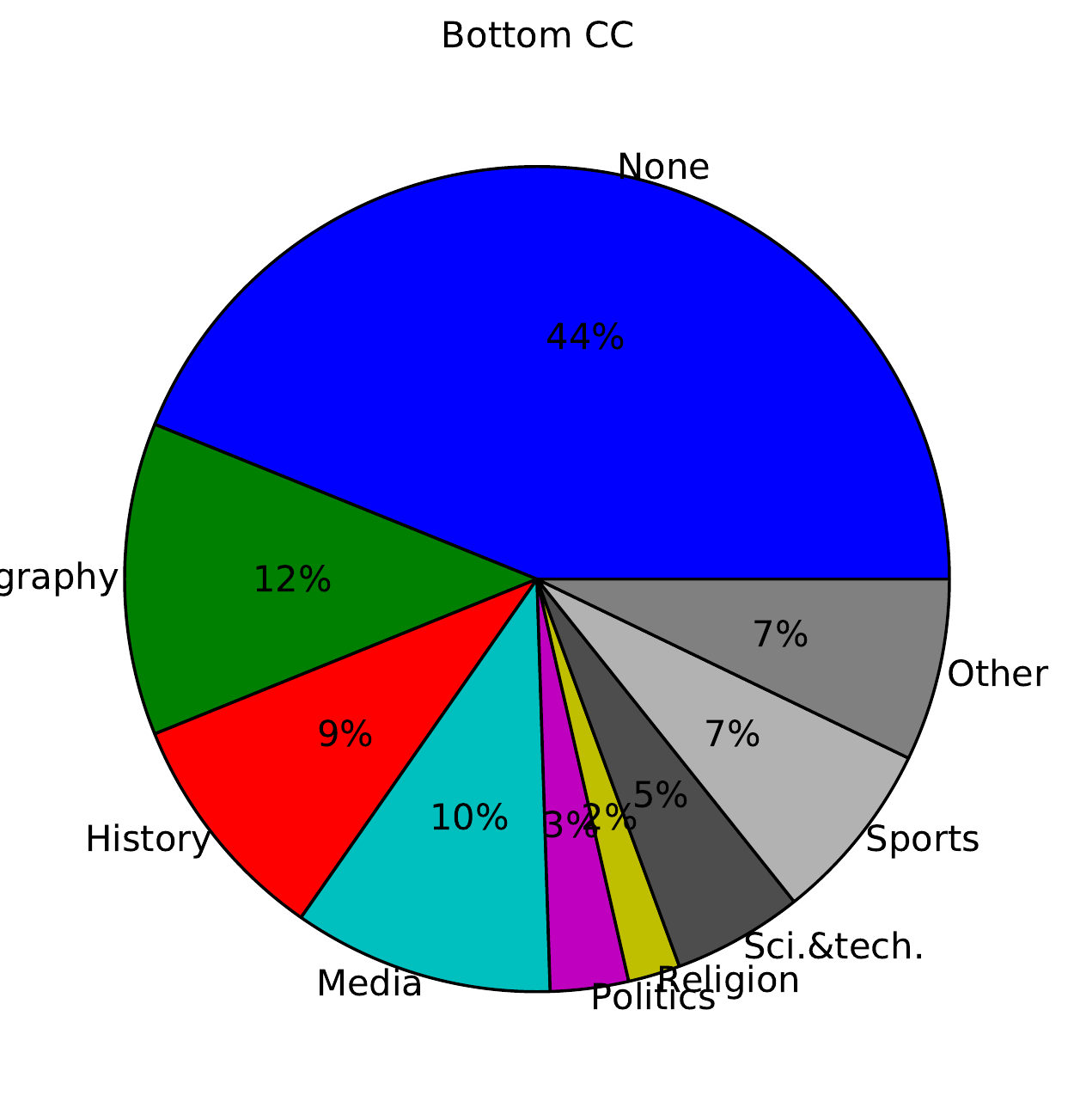}
  \end{center}
  \caption{Human evaluation of the general category of edits (if any)
    for administrators directly after their RfA. The 100 highest and
    100 lowest scoring administrators for each metric are shown;
    because of overlap, 286 administrators are represented in
    total. Many users with a high CC-Score are topically focused, and
    the CC-Score finds many more media-focused users than the
    C-Score.}
  \label{figure:human_eval}
\end{figure}

\subsection{Highest scoring admins}

We now turn to examining the behavior of admins through the lens of
the Controversy and CC-Scores. Where do admins focus their attention,
and what is controversial on Wikipedia? To explore this issue, we use
human evaluations of the edit history of administrators during the 180
days after they became an administrator. Without knowing anything
about the CC, Controversy, or Clustering Scores associated with the
edit history, a reviewer analyzed the top 50 pages edited by a user
and decided which general category, if any, the edits were in. Results
for the top and bottom 100 administrators ranked by each score are
presented in Figure \ref{figure:human_eval}.

Figure \ref{figure:human_eval} is useful validation for our
methodology: for example, administrators with very low CC-Scores were
often classified as not editing in a coherent general
category. Conversely, users with a high CC-Score were much more likely
to be topically focused. This effect is even more apparent for the raw
Clustering Score, with 73\% of the bottom 100 admins classified as
topically unfocused. This implies that our article similarity metric
corresponds with an intuitive notion of topical similarity.

There are some interesting differences in the topic areas of the 100
admins with the highest C-Scores and the 100 with the
highest CC-Scores. As expected, the C-Score picks up a
substantial chunk of editors with no particular focus, while the CC
score does not. These are admins who are doing their job of
``policing'' controversy across a broad spectrum of
topics. Surprisingly, the C-Score picks up more politically
focused editors while the CC-Score picks up many more who focus on
media and entertainment.

Specific examples help to elucidate this effect. Table
\ref{table:after_cc_controversy} shows the top 6 administrators by CC
and C-Score respectively for the 180 day period immediately
following their promotion (successful RfA).  The CC-Score picks up
three media-focused editors while the C-Score does
not. These editors focus on a specific media franchise (a TV show, for
example), editing on this topic almost exclusively. The long tail of
edits for media-focused users often includes non-controversial
articles related to the same media franchise, for example articles on
minor characters. This comprehensive long tail means that the
page-level clustering score for pages related to the media franchise
is very high, including the main pages related to the franchise; these
main pages tend to be quite controversial. This combination of
controversy and clustering contributes significantly to the user's
overall CC-Score, while the long tail of less controversial articles
moderates the C-Score.

We note that all the users with very high C-Scores also have very high
CC-Scores (falling at least in the 96th and typically in the 99th
percentile of CC-Score). However, the converse is not true, because of
the properties of the CC-Score described above. For example Admins 3,
4, and 5 in particular are relatively low in C-Score. Admin 5 is
media-focused. Admin 4 has a high CC-Score despite editing
topically unrelated pages; yet the similarity scores between them are
high. This user focuses on disambiguation pages, all of which
share common categories, and many of which are maintained by the same
users. Major disambiguation pages also turn out to be fairly
controversial, being of interest to editors contributing to any of the
pages they link to. This combination of a long tail of ``related''
disambiguation pages with controversial major disambiguation pages
leads to a high CC-Score, just as it did for media. As with media, the
CC-Score highlights a real phenomenon on Wikipedia: WikiProject
Disambiguation, of which this user is a member, consists of users who
focus their efforts on disambiguation pages. Thus the CC-Score is
finding exactly single-topic editors of controversial pages, even if
their topic is rather specific to Wikipedia.

The story with Admin 3 is similar to Admins 4 and 5, in that this
editor also edits a long tail of uncontroversial
articles. Qualitatively, it is useful that the CC-Score is picking up
different people than the simple C-Score, and sometimes
turning up surprises -- this is exactly the kind of behavior one
would want to pick up on, and it may give the CC-Score an advantage
over the simple C-Score in terms of detecting subtle
manipulation.

We note that in this paper we are agnostic to what makes a page or a
topic controversial, which reveals much of interest about Wikipedia,
but at the same time our methods are completely general --
specifically, we expect they would work well with any measure of
controversy, and so the techniques can easily be adapted to
domain-specific needs. For example, to focus on more traditionally
controversial single-topic editors, we might consider a modification
of the page-level controversy score which ignores controversy on
media-related pages. 
More generally, the page-level controversy score can give preference
to any topic; it does not need to be related to controversy at
all. For example, we could find single-topic editors interested in a
specific country.

\begin{table}
\begin{myfontsize}
\vspace{-.33in}
\begin{tabular}{p{.33\textwidth}p{.33\textwidth}p{.33\textwidth}}
\begin{tabular}[t]{p{.18\textwidth}rr} 
\multicolumn{3}{c}{\textbf{Admin1}}\\
\multicolumn{3}{c}{CC 100.0\% Clust. 96.4\% Cont. 99.9\%}\\ \hline
Title & Edit\% & Cont.\\ \hline Jehovah's Witnesses & 63.9\% & 100.0\%\\ Salt & 6.6\% & 99.7\%\\ Tony Blair & 1.6\% & 100.0\%\\ Joseph Franklin Rutherford & 1.6\% & 99.7\%\\ Robby Gordon & 1.6\% & 96.4\%\\ Leet & 1.6\% & 100.0\%\\ Jehovah's Witnesses and congregational discipline & 1.6\% & 99.5\%\\ Wavelength (disambiguation) & 1.6\% & 0.0\%\\ Charles Taze Russell & 1.6\% & 100.0\%\\ \end{tabular} &
\begin{tabular}[t]{p{.18\textwidth}rr} 
\multicolumn{3}{c}{\textbf{Admin2}}\\
\multicolumn{3}{c}{ CC 99.9\% Clust. 97.4\% Cont. 95.7\%}\\ \hline
Title & Edit\% & Cont.\\ \hline Sailor Venus & 4.2\% & 99.7\%\\ Sailor Moon (character) & 4.0\% & 99.9\%\\ Sailor Mercury & 3.7\% & 99.8\%\\ List of minor Sailor Moon characters & 3.1\% & 99.7\%\\ List of Sailor Moon episodes & 3.0\% & 99.8\%\\ The Church of Jesus Christ of Latter-day Saints & 2.7\% & 100.0\%\\ Sailor Starlights & 2.7\% & 99.6\%\\ Sailor Jupiter & 2.6\% & 99.6\%\\ Sailor Moon & 2.6\% & 100.0\%\\ \end{tabular} &
\begin{tabular}[t]{p{.18\textwidth}rr} 
\multicolumn{3}{c}{\textbf{Admin3}}\\
\multicolumn{3}{c}{ CC 99.9\% Clust. 98.8\% Cont. 76.6\%}\\ \hline
Title & Edit\% & Cont.\\ \hline Global city & 2.8\% & 99.9\%\\ List of
school pranks & 2.6\% & 99.7\%\\ Scholars for 9/11 Truth & 2.0\% &
99.9\%\\ Pi & 1.9\% & 100.0\%\\ 9/11 Truth movement & 1.8\% &
100.0\%\\ Christopher Langan & 1.8\% & 99.9\%\\ Hubbert peak theory &
1.8\% & 100.0\%\\ Stephen Barrett & 1.6\% & 100.0\%\\ Prime number &
1.2\% & 99.9\%\\ \end{tabular}\\ \hline
\begin{tabular}[t]{p{.18\textwidth}rr} 
\multicolumn{3}{c}{\textbf{Admin4}}\\
\multicolumn{3}{c}{ CC 99.8\% Clust. 99.9\% Cont. 37.9\%}\\ \hline
Title & Edit\% & Cont.\\ \hline Matrix & 5.4\% & 96.5\%\\ Apartheid (disambiguation) & 3.5\% & 99.8\%\\ ETA (disambiguation) & 3.5\% & 99.4\%\\ Gemini & 3.1\% & 92.8\%\\ Solo & 3.1\% & 59.0\%\\ XXX & 3.1\% & 96.8\%\\ X Games & 2.7\% & 96.9\%\\ FAST & 2.3\% & 39.2\%\\ Fame & 2.3\% & 74.7\%\\ \end{tabular} &
\begin{tabular}[t]{p{.18\textwidth}rr} 
\multicolumn{3}{c}{\textbf{Admin5}}\\
\multicolumn{3}{c}{ CC 99.8\% Clust. 99.3\% Cont. 72.3\%}\\ \hline
 Title & Edit\% & Cont.\\ \hline Meet Kevin Johnson & 14.7\% & 98.8\%\\ The Other Woman & 10.1\% & 97.8\%\\ Lost (season 4) & 7.2\% & 99.6\%\\ Lost (season 5) & 6.8\% & 99.3\%\\ List of Heroes episodes & 6.2\% & 99.9\%\\ Martin Keamy & 5.3\% & 96.9\%\\ The Shape of Things to Come (Lost) & 4.0\% & 98.3\%\\ Through the Looking Glass (Lost) & 1.8\% & 99.6\%\\ There's No Place Like Home & 1.7\% & 98.0\%\\ \end{tabular} &
\begin{tabular}[t]{p{.18\textwidth}rr} 
\multicolumn{3}{c}{\textbf{Admin6}}\\
\multicolumn{3}{c}{ CC 99.7\% Clust. 95.1\% Cont. 83.5\%}\\ \hline
Title & Edit\% & Cont.\\ \hline Big Brother 2007 (UK) & 17.6\% & 100.0\%\\ List of Big Brother 2007 housemates (UK) & 6.7\% & 98.9\%\\ Ionic bond & 6.1\% & 98.4\%\\ Big Brother (UK) & 3.0\% & 99.9\%\\ Big Brother 8 (U.S.) & 2.4\% & 100.0\%\\ Big Brother X & 2.4\% & 38.8\%\\ Big Brother 2006 (UK) & 2.4\% & 100.0\%\\ Big Brother (TV series) & 2.4\% & 99.8\%\\ Big Brother 2004 (UK) & 2.4\% & 99.6\%\\ \end{tabular} \\ \hline \hline

\begin{tabular}[t]{p{.18\textwidth}rr}  
\multicolumn{3}{c}{\textbf{Admin7}}\\
\multicolumn{3}{c}{CC 100.0\% Clust. 96.4\% Cont. 99.9\%}\\ \hline
Title & Edit\% & Cont.\\ \hline Jehovah's Witnesses & 63.9\% & 100.0\%\\ Salt & 6.6\% & 99.7\%\\ Tony Blair & 1.6\% & 100.0\%\\ Joseph Franklin Rutherford & 1.6\% & 99.7\%\\ Robby Gordon & 1.6\% & 96.4\%\\ Leet & 1.6\% & 100.0\%\\ Jehovah's Witnesses and congregational discipline & 1.6\% & 99.5\%\\ Wavelength (disambiguation) & 1.6\% & 0.0\%\\ Charles Taze Russell & 1.6\% & 100.0\%\\ \end{tabular} &
\begin{tabular}[t]{p{.18\textwidth}rr} 
\multicolumn{3}{c}{\textbf{Admin8}}\\
\multicolumn{3}{c}{ CC 99.6\% Clust. 89.6\% Cont. 99.8\%}\\ \hline
Title & Edit\% & Cont.\\ \hline Global warming & 9.7\% & 100.0\%\\ Global warming controversy & 5.6\% & 100.0\%\\ Electronic voice phenomenon & 3.6\% & 100.0\%\\ An Inconvenient Truth & 3.3\% & 100.0\%\\ Greenhouse effect & 3.0\% & 100.0\%\\ American Enterprise Institute & 2.9\% & 99.8\%\\ The Great Global Warming Swindle & 2.8\% & 100.0\%\\ List scientists opposing the mainstream scientific assessment of global warming & 2.8\% & 100.0\%\\ The Beatles & 2.6\% & 100.0\%\\ \end{tabular} &
\begin{tabular}[t]{p{.18\textwidth}rr} 
\multicolumn{3}{c}{\textbf{Admin9}}\\
\multicolumn{3}{c}{ CC 99.5\% Clust. 87.5\% Cont. 99.8\%}\\ \hline
Title & Edit\% & Cont.\\ \hline 1948 Palestinian exodus & 6.8\% & 100.0\%\\ Yasser Arafat & 6.3\% & 100.0\%\\ Israeli West Bank barrier & 4.9\% & 100.0\%\\ Israeli settlement & 4.9\% & 100.0\%\\ Hebron & 4.6\% & 100.0\%\\ Second Intifada & 3.2\% & 100.0\%\\ Gaza Strip & 3.2\% & 99.9\%\\ Palestinian territories & 2.9\% & 100.0\%\\ Palestinian people & 2.9\% & 100.0\%\\ \end{tabular} \\ \hline
\begin{tabular}[t]{p{.18\textwidth}rr} 
\multicolumn{3}{c}{\textbf{Admin10}}\\
\multicolumn{3}{c}{ CC 96.3\% Clust. 52.1\% Cont. 99.7\%}\\ \hline
Title & Edit\% & Cont.\\ \hline Rick Reilly & 10.0\% & 99.6\%\\ Keith Olbermann & 6.4\% & 100.0\%\\ Lara Logan & 5.5\% & 99.9\%\\ Treaty of Tripoli & 4.5\% & 99.9\%\\ Glenn Greenwald & 3.6\% & 99.9\%\\ Eli Whitney, Jr. & 3.6\% & 99.3\%\\ Michael J. Fox & 3.6\% & 99.8\%\\ Newton's laws of motion & 3.6\% & 99.9\%\\ William Connolley & 2.7\% & 100.0\%\\ \end{tabular} &
\begin{tabular}[t]{p{.18\textwidth}rr} 
\multicolumn{3}{c}{\textbf{Admin11}}\\
\multicolumn{3}{c}{ CC 96.9\% Clust. 59.6\% Cont. 99.6\%}\\ \hline
Title & Edit\% & Cont.\\ \hline Abortion & 36.6\% & 100.0\%\\ University of Michigan & 5.5\% & 99.9\%\\ Jesus & 3.4\% & 100.0\%\\ Islamofascism & 2.8\% & 100.0\%\\ NARAL Pro-Choice America & 2.8\% & 99.0\%\\ Intelligent design & 2.8\% & 100.0\%\\ Saint Joseph & 2.1\% & 99.9\%\\ C. S. Lewis & 2.1\% & 100.0\%\\ God & 1.4\% & 100.0\%\\ \end{tabular} &
\begin{tabular}[t]{p{.18\textwidth}rr} 
\multicolumn{3}{c}{\textbf{Admin12}}\\
\multicolumn{3}{c}{ CC 97.7\% Clust. 69.3\% Cont. 99.6\%}\\ \hline
Title & Edit\% & Cont.\\ \hline Yasser Arafat & 5.2\% & 100.0\%\\ Estimates of the Palestinian Refugee flight of 1948 & 2.7\% & 99.6\%\\ Israel & 2.2\% & 100.0\%\\ ArabIsraeli conflict & 2.0\% & 100.0\%\\ Ariel Sharon & 1.8\% & 100.0\%\\ Jews & 1.8\% & 100.0\%\\ Antisemitism & 1.5\% & 100.0\%\\ Muhammad al-Durrah incident & 1.5\% & 100.0\%\\ Second Intifada & 1.5\% & 100.0\%\\ \end{tabular}\\
\end{tabular}
\caption{The most edited articles by the administrators with the
  highest CC-Scores (top 6) and highest C-Scores (bottom 6)
  during the 180 days after they became an admin. Each article is
  annotated with the percentile of its article-level controversy score
  and the percentage of the administrator's edits which were to that
  article. On top of each table are the percentiles for the CC,
  Clustering, and C-Scores of the administrator during the
  same period.}
\label{table:after_cc_controversy}
\end{myfontsize}
\end{table}

\subsection{Distribution of CC changes}
While it is interesting to find editors focused on a single,
controversial topic, it is not surprising that such editors exist;
Wikipedia certainly needs domain experts even on controversial
topics. Sudden changes in behavior, especially increases in the
topical concentration or the controversial nature of edits, are more
surprising; especially so when some level of community trust is
involved, as with administrators. In particular, an editor changing
behavior dramatically shortly after becoming an admin is suspicious.

\paragraph{The RfA process} Standing for promotion to adminship on
Wikipedia is an involved process.\footnote{%
  \url{http://en.wikipedia.org/wiki/Wikipedia:Rfa} } %
An editor who stands for, or is nominated for adminship must undergo a
week of public scrutiny which allows the community to build consensus
about whether or not the candidate should be promoted. A special page
is set up on which the candidate makes a nomination statement about
why she or he should be promoted, based on detailed evidence from
their history of contributions to Wikipedia. Other users can then
weigh in and comment on the case, and typically a large volume of
support (above 75\% of commenters) as well as solid supporting
statements from other editors are necessary for high-level Wikipedia
``bureaucrats'' to approve the application. Burke and Kraut provide
many further details on this process \cite{Burke2008Mop}. Wikipedia
policies call for nominees to demonstrate a strong edit history,
varied experience, adherence to Wikipedia policies on points of view
and consensus, as well as demonstration of willingness to help with
tasks that admins are expected to do, like building consensus. Burke
and Kraut note that the actual value of some of these may be mixed:
participating in seemingly controversial tasks like fighting vandalism
or requesting admin intervention on a page before becoming an admin
actually seems to hurt the chances of success. 

Overall, the Wikipedia community devotes significant effort to the RfA
process, and there is a lot of human attention focused on making sure
that those who become admins are worthy of the community's trust. Now
we turn to examining some cases where the behavior of an editor
changed significantly right after they became an admin.

\begin{table}[t]
\begin{myfontsize}
\begin{tabular}{@{}c@{}c}
\begin{tabular}[t]{c@{}c}
\multicolumn{2}{c}{\textbf{Admin 1: Rank 1}}\\ \hline
\textbf{Before RfA} & \textbf{After RfA}\\
\begin{tabular}[t]{p{.15\textwidth}r}
Article & Edit\%\\\hline
Jehovah's Witnesses & 48.5\%\\
Eschatology of Jehovah's Witnesses & 3.8\%\\
Jehovah's Witnesses practices & 2.1\%\\
Organizational structure of Jehovah's Witnesses & 1.6\%\\
Criticism of Jehovah's Witnesses & 1.5\%\\
History of Jehovah's Witnesses & 1.4\%\\
Christianity & 1.3\%\\
Controversies regarding Jehovah's Witnesses & 1.1\%\\
Beliefs and practices of Jehovah's Witnesses & 0.8\%\\
\end{tabular} & 
\begin{tabular}[t]{p{.15\textwidth}r}
Article & Edit\%\\\hline
Jehovah's Witnesses & 62.1\%\\
Jehovah's Witnesses and the United Nations & 6.8\%\\
Criticism of Jehovah's Witnesses & 4.9\%\\
Salt & 3.9\%\\
Jehovah's Witnesses and congregational discipline & 2.9\%\\
Eschatology of Jehovah's Witnesses & 1.9\%\\
Blue link & 1.0\%\\
Brendan Loy & 1.0\%\\
Charles Taze Russell & 1.0\%\\
\end{tabular}\\
\end{tabular}
 & 
\begin{tabular}[t]{c@{}c}
\multicolumn{2}{c}{\textbf{Admin 2: Rank 3}}\\ \hline
\textbf{Before RfA} & \textbf{After RfA}\\
\begin{tabular}[t]{p{.15\textwidth}r}
Article & Edit\%\\\hline
Chiropractic & 6.5\%\\
Extreme physical information & 3.6\%\\
Prime number & 3.4\%\\
Normal number & 3.2\%\\
Axiom of choice & 3.1\%\\
Year 10,000 problem & 2.2\%\\
Difference operator & 2.0\%\\
Kenny Rogers Roasters & 1.7\%\\
Selector calculus & 1.7\%\\
\end{tabular} & 
\begin{tabular}[t]{p{.15\textwidth}r}
Article & Edit\%\\\hline
Global city & 3.9\%\\
Christopher Langan & 2.3\%\\
Stephen Barrett & 2.0\%\\
Scholars for 9/11 Truth & 1.9\%\\
List of school pranks & 1.8\%\\
The National Council Against Health Fraud & 1.7\%\\
9/11 Truth movement & 1.6\%\\
Quackwatch & 1.6\%\\
Pi & 1.5\%\\
\end{tabular}\\
\end{tabular}
\\ \hline
\begin{tabular}[t]{c@{}c}
\multicolumn{2}{c}{\textbf{Admin 3: Rank 5}}\\ \hline
\textbf{Before RfA} & \textbf{After RfA}\\
\begin{tabular}[t]{p{.15\textwidth}r}
Article & Edit\%\\\hline
Rechargeable battery & 8.0\%\\
Flywheel energy storage & 6.5\%\\
Ethanol fuel & 3.0\%\\
Imaginary color & 3.0\%\\
Mensural notation & 2.7\%\\
Noise pollution & 2.5\%\\
Pay it forward & 2.1\%\\
TamilNet & 1.8\%\\
CIE 1931 color space & 1.6\%\\
\end{tabular} & 
\begin{tabular}[t]{p{.15\textwidth}r}
Article & Edit\%\\\hline
Buddhahood & 13.2\%\\
Sri Lanka and state terrorism & 3.9\%\\
Premier of the Republic of China & 3.0\%\\
Sri Lankan Tamil militant groups & 2.8\%\\
Outpost for Hope & 2.6\%\\
Sri Lanka Armed Forces & 2.6\%\\
Liberation Tigers of Tamil Eelam & 2.1\%\\
Esperanto & 1.4\%\\
List of attacks attributed to the LTTE & 1.4\%\\
\end{tabular}\\
\end{tabular}
 & 
\begin{tabular}[t]{c@{}c}
\multicolumn{2}{c}{\textbf{Admin 4: Rank 9}}\\ \hline
\textbf{Before RfA} & \textbf{After RfA}\\
\begin{tabular}[t]{p{.15\textwidth}r}
Article & Edit\%\\\hline
Biman Bangladesh Airlines & 8.4\%\\
Fatimah & 6.3\%\\
First Solution Money Transfer & 3.0\%\\
72 Virgins & 1.8\%\\
2007 Bangladesh cartoon controversy & 1.3\%\\
Ramadan & 1.2\%\\
Royal Bengal Airline & 1.2\%\\
Criticism of the Qur'an & 1.2\%\\
Air Sylhet & 1.1\%\\
\end{tabular} & 
\begin{tabular}[t]{p{.15\textwidth}r}
Article & Edit\%\\\hline
Mawlid & 20.0\%\\
Fatimah & 11.8\%\\
Five Pillars of Islam & 3.6\%\\
Air Sylhet & 3.0\%\\
Ezra & 2.5\%\\
Hajj & 2.1\%\\
Osmani International Airport & 2.0\%\\
Biman Bangladesh Airlines & 1.8\%\\
Eid ul-Fitr & 1.8\%\\
\end{tabular}\\
\end{tabular}
\\
\end{tabular}
\caption{Most edited articles for 180 days before and after becoming
  an administrator. Users were selected from the top ten CC-Score
  changes.}
\label{table:cc_changes}
\end{myfontsize}
\end{table}

\paragraph{Changes in behavior}
Table \ref{table:cc_changes} shows the article edit history of four
administrators for 180 days before and 180 days after their successful
RfA. These users were among the top 10 administrators ranked by the
change in CC-Score between the two periods (Admins 1, 2, 3, and 4 were
ranked 1, 3, 5, and 9 respectively in CC-Score change). For two of the
administrators shown (Admin 1 and Admin 4), the change in CC-Score is
explained by a change in edit distribution; while the general category
of their edits remains consistent, they concentrated on this category
much more heavily after their respective RfAs.

For Admins 2 and 3 in Table \ref{table:cc_changes}, the change in CC
Score is the result of a rather dramatic shift in topic. Admin 2
shifts from mathematics to 9/11 conspiracy theories (several related
pages are not shown in the table), while Admin 3 shifts from
relatively unfocused edits to the Sri Lankan Civil War. Upon further
examination of their behavior, neither administrator appears to be
violating Wikipedia policy 
(instead acting as mediators and enforcing a neutral point of view),
and yet the changes are quite striking. While it may not be the case
for these editors, a similar pattern could reflect subtle manipulation
by one-sided enforcement of the NPOV guidelines, for example.

\begin{figure*}[t]
  \subfigure[Full distribution] {
    \includegraphics[width=.5\textwidth]{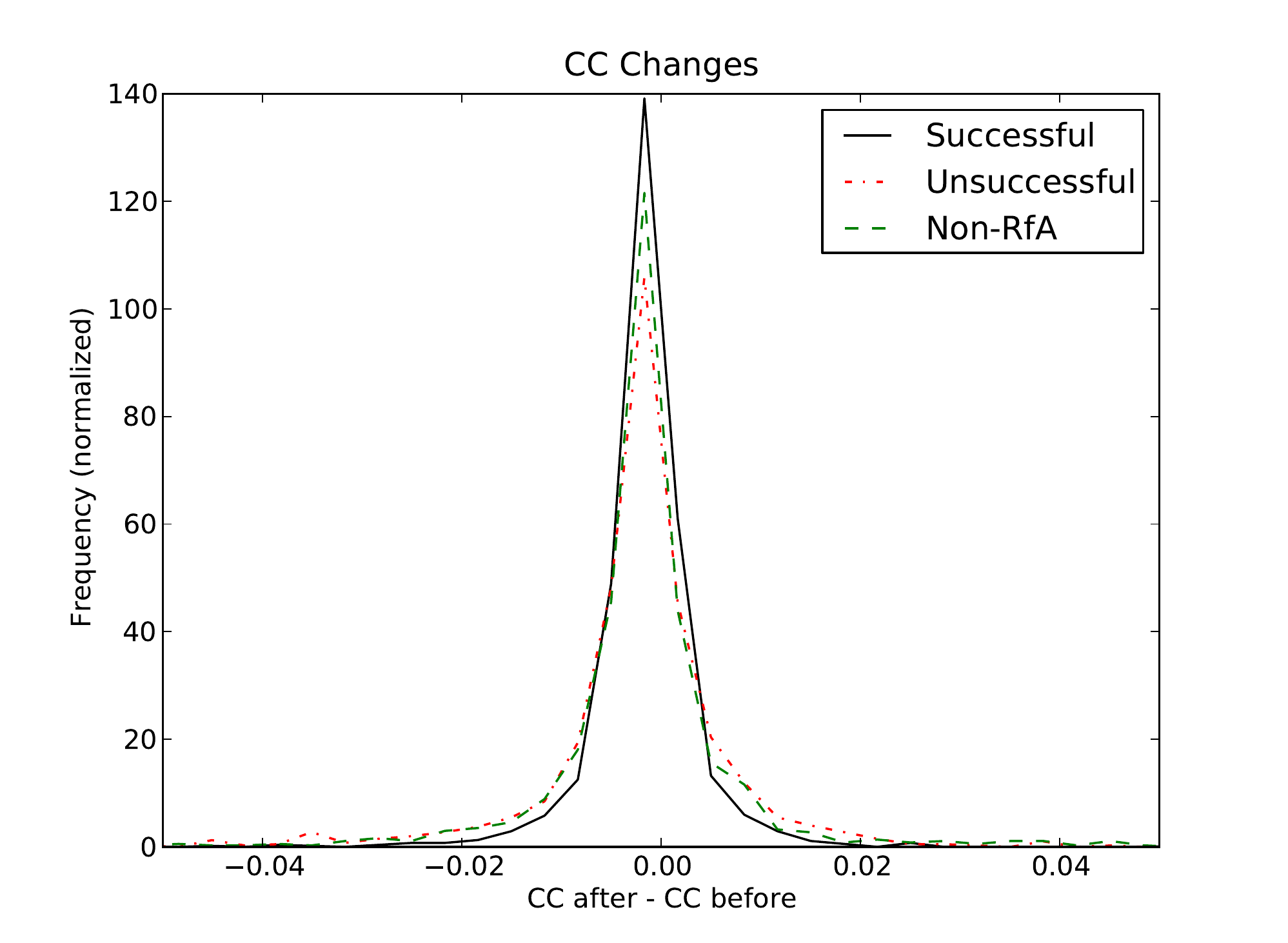}
  }
  \subfigure[Log-log plot of the right hand tail] { 
    \includegraphics[width=.5\textwidth]{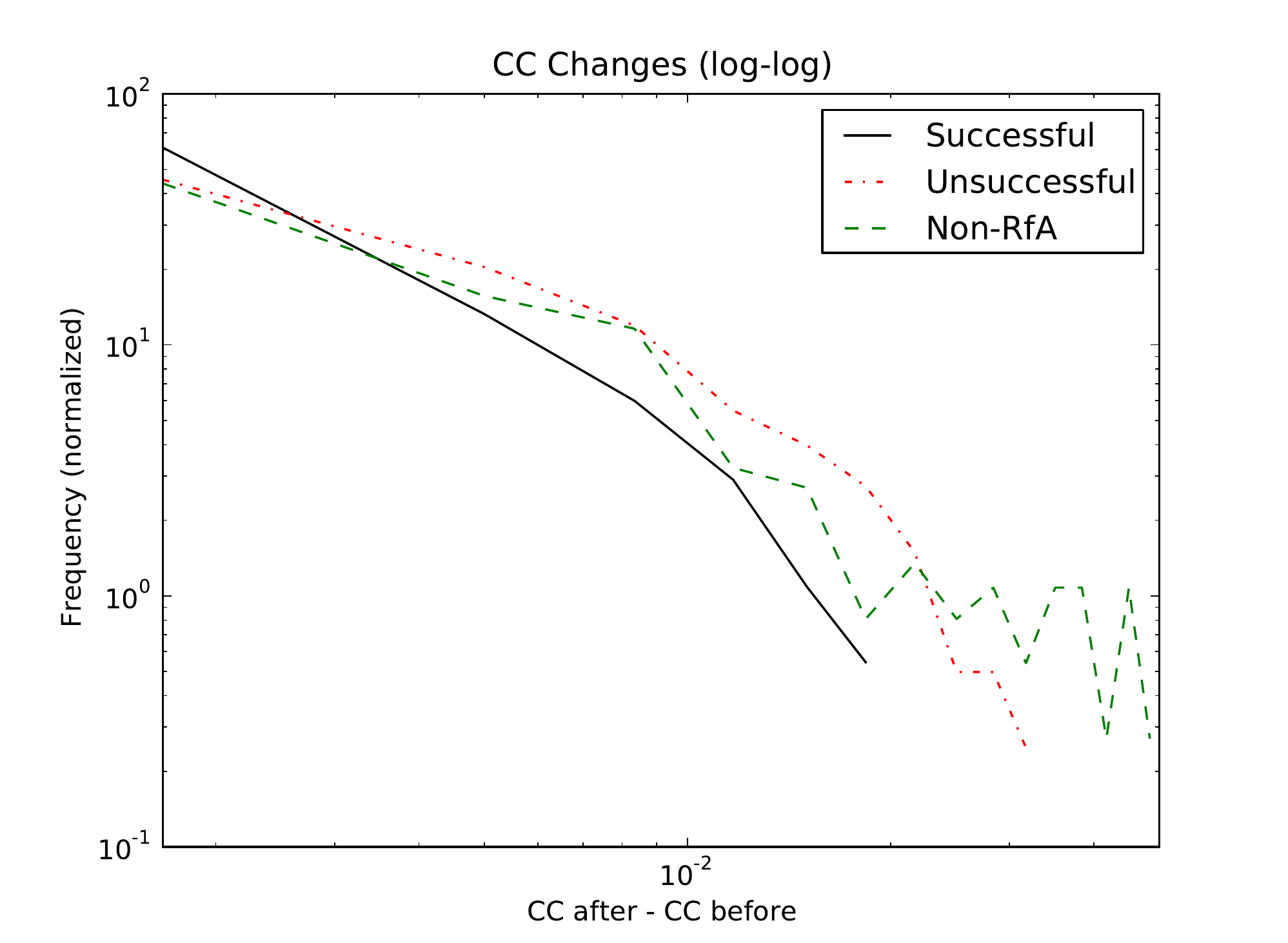}
  }
  \caption{Distribution of changes in the CC-Score
    before and after successful and unsuccessful RfAs, and for users
    who have never participated in an RfA.}
  \label{figure:cc_change_dist}
\end{figure*}

\paragraph{Population-level changes}
This leads us to a more general question. Is there evidence of a
population-wide change in Admin behavior after successful RfAs?
Figure \ref{figure:cc_change_dist} shows the distribution of CC-Score
changes for RfA candidates, both successful and unsuccessful, before
and after their respective RfAs (again 180 days each), and for a group
of 1000 active editors who were never nominated for administrator
status. The results clearly show that those who stand for promotion
and are successful behave differently at the population-level than
those who either stand for promotion and fail or those who never stand
for promotion at all. In fact, they end up staying closer to their
previous behavior than either of the other groups -- the variance in
CC-Score changes is higher for the other two groups than it is for
editors who had successful RfAs (see below for details on the data and
statistical tests). This implies strongly that there is no serious
problem with people becoming admins on Wikipedia in order to push
their own point of view. There are two reasonable hypotheses that may
explain the lower variance in CC-Score changes for successful RfA
candidates. Either some aspect of the RfA process selects for editors
who are less likely to change their behavior, or the very fact of
becoming an administrator has a ``centralizing'' influence: given
their new status, associated with a (real or perceived) higher level
of scrutiny, administrators become less likely to change their
behavior.

The hypothesis that the RfA process selects for editors who tend not
to change their CC-Scores is unlikely, as we would then expect this
type of user to appear in the population of users who were never
nominated to become administrators. If this were the case, then the
non-RfA distribution would be a mix of the successful and unsuccessful
RfA distributions; instead, the unsuccessful and non-RfA distributions
are similar to each other and different from successful RfAs. We run a
second test that provides further evidence for the centralizing
hypothesis. We construct a matched sample of successful and unsuccessful
RfA candidates, matching on the \emph{estimated probability} $p_i$
that editor $i$'s RfA will be successful RfA based on $i$'s pre-RfA
behavior. We use the model of Burke and Kraut to estimate the $p_i$s
\cite{Burke2008Mop}. For each successful RfA $j$, we find the editor
in the unsuccessful RfA set $k$ with $p_k$ closest to $p_j$ (throwing
out examples that are not within 1 percentage point). We then compare
the population-level behavior of the two sets of editors we are left
with. Now that we have controlled for endogenous factors, we expect
that the two populations are very similar in intrinsic qualities: the
only difference between them should be that the successful ones
actually became admins and the unsuccessful ones did not. We again
find that the population of successful admins is significantly
different, exhibiting more stable behavior than the population of
editors who were unsuccessful in their RfAs (see below for details on
statistics).  This suggests that some aspect of actually being an
administrator reduces the propensity for significant behavior changes
(incidentally, this makes administrators who do significantly change
their editing behavior all the more interesting).

\textbf{Data and statistics: }
For the non-RfA users, we use edits before and after a randomly
selected edit. The distributions of changes for unsuccessful and
non-RfA editors have a significantly higher variance than the
distribution for successful RfAs, with the 95\% confidence interval on
the ratio of the variance of the successful RfA distribution to the
variance among non-RfA users being $[0.22, 0.28]$. The 95\% confidence
interval on the same ratio for unsuccessful and non-RfA users, on the
other hand, is $[0.88, 1.11]$ (est. $0.99$). Further, the
Kolmogorov-Smirnov Test rules out an identical distribution for
successful and unsuccessful distributions ($p=10^{-7}$, $D=0.109$),
and between the successful and non-RfA distributions ($p=10^{-4}$,
$D=0.086$). We cannot rule out the possibility that the non-RfA and
unsuccessful distributions are identical ($p=0.25$, $D=0.042$). Closer
examination of the tails of the distributions does not show any
differences not already explained by the variance. For the matched
sample described above, the 95\% confidence interval for the ratio of
the variances of successful and unsuccessful RfA distributions is
[0.32, 0.43]; the conclusions of the KS-tests are unchanged.

\section{Discussion}
\label{sec:discussion}
This paper contributes to the literature in two different ways: first,
we introduce new behavioral metrics for quantifying controversial
editing on Wikipedia. The measures we introduce can be used (perhaps
with domain specific modifications) to triage suspicious behavior for
deeper investigation. Second, these measures allow us to contribute to
the study of Wikipedia as an evolving social system: along with
showing that Wikipedia admins behave in a stable manner, we also
identify some intuitively surprising topics of conflict in Wikipedia.

The Controversy Score (C-Score), measures the extent to which an
editor is influencing controversial pages. The Clustered Controversy
Score (CC-Score) builds on the C-Score, finding single-topic editors
of controversial pages. Both metrics are flexible, since they are
language- and platform- independent, and can work with different
measures of controversy.

We validate the C- and CC-Scores as user-level measures of controversy
on a set of blocked users. On a set of administrators, we find several
weaknesses of the C-Score: it misses controversial, single-topic
editors in the presence of a long tail of related, but less
controversial, pages. Further, the C-Score gives undue weight to
unfocused vandalism fighting, a common behavior among
administrators. The CC-Score solves many of these issues, allowing us
to find controversial, single-topic editors, and often finds editors
who would not show up ranked highly on just the C-Score measure. The
CC-score also enables us to identify interesting Wikipedia-specific
phenomena, for example, the substantial levels of controversy
associated with some media/entertainment specific pages as well as
with some disambiguation pages.

We also show how the CC-Score can be used to analyze behavior changes,
both for single editors and in aggregate. We find several instances of
dramatic shifts in behavior by administrators upon assuming their
responsibilities. At the same time, we show that administrators as a
group change their behavior significantly less than any other group of
Wikipedians. This consistency appears to be due to the role of
administrator itself, rather than being a selection effect.

\paragraph{Future work} While we focus on editors working alone in
this paper, an extension of the CC-Score might highlight groups of
editors influencing a single, controversial topic; this presents
interesting computational and evaluation challenges. Improvements to
the CC-Score to better detect manipulation might focus on natural
language processing, or on non-local aspects of an editor's
behavior. Being platform-independent, the CC-Score is a useful tool
for analyzing behavior in general collective wisdom processes; we are
interested in applications of the CC-Score to other domains.

\section{Acknowledgments}
The authors would like to thank Mark Goldberg and Al Wallace for helpful
discussions.
This research is supported by an NSF CAREER Award (IIS-0952918) to
Das. Magdon-Ismail's contributions to 
this research are continuing through participation in the 
Network Science Collaborative Technology Alliance sponsored 
by the U.S. Army Research Laboratory under
Agreement Number W9\-11NF-09-2-0053. The views and
conclusions contained in this document are those of the 
authors and should not be 
interpreted as representing the official policies, either
expressed or implied, of the Army Research Laboratory or the U.S. Government.

\bibliographystyle{abbrv}
\bibliography{wiki}
\end{document}